\documentclass[aps,pra,groupedaddress,twocolumn,showpacs]{revtex4}

\usepackage{graphicx,subfigure}
\usepackage{amsmath}

\newcommand{\ket}[1]{|#1\rangle}

\newcommand{\eq}{\begin{equation}}
\newcommand{\fine}{\end{equation}}

\newcommand{\ii}{\text i}
\newcommand{\bc}{\begin{cases}\begin{aligned}}
\newcommand{\ec}{\end{aligned}\end{cases}}
\newcommand{\z}{x}

\begin{document}
%\preprint{}
\title{Experimental Realization of Polarization Qutrits from Non-Maximally Entangled States}


\author{Giuseppe Vallone}
\homepage[]{http://quantumoptics.phys.uniroma1.it/}

\author{Enrico Pomarico}
\homepage[]{http://quantumoptics.phys.uniroma1.it/}

\author{Paolo Mataloni}
\homepage[]{http://quantumoptics.phys.uniroma1.it/}

\author{Francesco De Martini}
\homepage[]{http://quantumoptics.phys.uniroma1.it/}
\affiliation{Dipartimento di Fisica dell'Universit\`{a} ``La Sapienza'' and
Consorzio Nazionale Interuniversitario per le Scienze Fisiche della Materia,
Roma, 00185 Italy}

\author{Marco Barbieri}
\affiliation{Dipartimento di Fisica dell'Universit\`{a} ``La Sapienza'', Roma, 00185 Italy and\\
Centre for Quantum Computer Technology, Department of Physics, University of Queensland, QLD 4072, Brisbane, Australia.}

\date{\today}

\begin{abstract}
Based on a recent proposal [Phys. Rev. A, {\bf 71}, 062337 (2005)], we have experimentally realized two photon 
polarization qutrits by using non-maximally entangled states and linear optical transformations. By this technique
high fidelity mutually unbiased qutrits are generated at a high brilliance level.
\end{abstract}


\maketitle

\section{Introduction}

Claude Shannon elected the bit as the fundamental unit of information. 
A system which can be only ``on'' or ``off'' is the simplest choice, but no fundamental reason prevent to adopt 
 $d>2$ logical levels for information processing. Nowadays, qudits, i.e. $d$ level quantum systems, can be easily engineered, 
 controlled and measured, 
 thus ensuring more freedom in choosing which dimensionality to use. The interest for these systems resides on the fact that dealing with 
 arbitrary dimensions may allow to  simplify the general structure of a quantum protocol. Moreover, quantum key distribution 
 schemes have been demonstrated to be more resilient to a specific class of eavesdropping attacks when adopting qutrits ($d=3$) or ququads ($d=4$) 
 instead of qubits \cite{00-bec-qua, 02-bru-opt, 02-cer-sec, 03-dur-sec}. 
 Multi-level systems and in particular qutrits are shown to be more efficient 
 also for designing other security protocols, e.g. bit commitment or coin tossing \cite{04-lan-mea,05-mol-exp}, 
 and for fundamental tests of quantum mechanics \cite{04-the-bel,02-col-bel,02-kas-cla}. 

Some optical realizations and applications of qutrits, exploiting different physical processes, have been demonstrated \cite{05-bar-gen}.
Time bin entangled qudits are generated by a time-frequency entangled photon pair through a multi-armed Franson interferometer \cite{04-the-bel}. 
In this case the dimensionality $d$ is given by the number of arms. This scheme presents a certain rigidity in switching among different states. 
A different approach exploits orbital angular momentum entanglement of single photons generated by Spontaneous Parametric Down Conversion
(SPDC), but only a partial control of the qutrit state is provided. Indeed, in the method of Refs. \cite{04-lan-mea,05-bar-gen,04-mol-tri,06-gro-exp}
a specific hologram is needed for each qutrit state. Transverse momentum correlation has also been used
to realize spatial bins \cite{05-osu-pix,05-nev-gen}. 
However, also in this case it seems unclear how to perform efficiently the rotation of the generated state. 

More recently, the experimental realization of arbitrary qutrit states, adopting the polarization degree of freedom of a two-photon state, 
was reported \cite{04-bog-qut}. By this technique three parametric sources, two type I and one type II nonlinear crystals,
placed respectively within and outside an interferometer, are shined by a common laser and determines
the critical adjustment of the qutrit phase. Moreover, the two collinear photons determining the qutrit state
are divided by a symmetric BS. This contributes to further
reduce the quite low production rate of the 3-level systems.

It is worth noting that qutrits have also been 
prepared by postselection from a four photon entangled state \cite{02-how-exp}.

In this paper we present the experimental realization of the proposal of Ref. \cite{05-dar-gen} to generate qutrits by using a single
non linear crystal and linear optical elements such as wave plates. Qutrits are encoded in the polarization of two photons 
initially prepared in a non-maximally entangled state, which plays the role of a ``seed'' state. 
Mutually unbiased bases can be obtained by linear 
optical transformations acting on two different seeds. This technique presents the advantage 
of merging accurate control and flexibility in the generation of the state at a high brilliance level.

The paper is organized as follows. 
Section \ref{sec:theory} concerns the description of the theoretical proposal of \cite{05-dar-gen}.
We explain how to generate a two photon polarization qutrit starting from a non-maximally entangled state and using
linear optics elements.
Section \ref{sec:experiment} shows the experimental results obtained by our technique. 
First we describe the source of entangled photons used in our experiment (subsection \ref{sec:source})
and present the experimental realization of the seed states (\ref{sec:seed}).
Then, in subsection \ref{sec:hada} and \ref{sec:phase}, the last stage of qutrits preparation, namely
the application of unitary transformations to each photon, is shown.

\section{\label{sec:theory}Theory}
Let's consider the polarization qutrit
\eq\label{xi}
\ket{\xi_{\psi,\phi}}=\frac{1}{\sqrt3}\left(\ket{H}_1\ket H_2+e^{\ii\psi}\ket{V}_1\ket V_2+
e^{\ii\phi}\ket{\psi^+}_{12}\right)\,,
\fine
where $1$ and $2$ label the two particles, $|H\rangle $ and $|V\rangle $ correspond to the horizontal and
vertical polarization states and $\ket{\psi^+}_{12}=\frac{1}{\sqrt{2}}(\ket{H}_1\ket{V}_2+\ket{V}_1\ket{H}_2)$ 
is one of the four polarization Bell states. The states in eq. \eqref{xi} span the symmetrical subspace of the two qubits Hilbert space. 

We are interested to the generation of a set of mutually unbiased (m.u.) bases, which are the basic tool for quantum key distribution
\cite{00-bec-qua,84-ben-qua}. 
On this purpose, we require that in the superposition state \eqref{xi}, 
the three terms of the computational basis $\{\ket{H}_1\ket{H}_2, \ket{V}_1\ket{V}_2, \ket{\psi^+}_{12}\}$ appear 
with the same probability amplitude. 
Indeed, our method is suitable to adjust at the same time both the balancement between 
the three contributions and the phases $\phi$ and $\psi$ needed to obtain m.u. bases.

Such states are obtained by applying two unitaries to a {\it seed} non-maximally entangled state,
\begin{equation}
\ket{\chi_{\psi,\phi}}=d_H\ket{H}_1\ket H_2+d_V\ket{V}_1\ket V_2\,.
\end{equation}
The dependence on the phases $\psi$ and $\phi$ is implicit in $d_H$ and $d_V$, which are chosen to be  real numbers:
\eq\label{x+-}
d_{H}=|\z_+|\,,\quad d_{V}=|\z_-|\,,
\fine
where
\eq
\z_\pm=\frac{\sqrt2\pm e^{\ii(\phi-\frac\psi2)}}{\sqrt6}\,.
\fine
 
We can write explicitly the transformation which maps the seed state $\ket{\chi_{\psi,\phi}}$
into the desired qutrit state as
\eq\label{xi-chi}
\ket{\xi_{\psi,\phi}}=(U\otimes W)\ket{\chi_{\psi,\phi}}\,,
\fine
up to an irrelevant global phase. 
The two unitaries $U$ and $W$, applied to photons $1$ and $2$ respectively and
expressed in the $\{\ket H,\ket V\}$ basis, are:
\begin{align}\label{W}
&W=
\underbrace{
\begin{pmatrix}
1 & 0\\
0 & e^{\ii\alpha}
\end{pmatrix}}_{P_{\alpha}}
\underbrace{\frac1{\sqrt2}
\begin{pmatrix}
1 & 1\\
-1 & 1
\end{pmatrix}}_{H'},\quad&&\alpha=\frac\psi2+\pi\\
\label{U}
&U=W
\begin{pmatrix}
1 & 0\\
0 & e^{\ii\Gamma}
\end{pmatrix}\,,&&\Gamma=\text{arg}\left(\frac{\z_-}{\z_+}\right)\,.
\end{align}
The phase shift $\Gamma$ can be introduced contextually with the generation of the seed state.
Indeed, thanks to the explicit expression of $U$ and $W$, the eq. \eqref{xi} can be written as
\eq\label{xi'}
\ket{\xi_{\psi,\phi}}=(P_\alpha\otimes P_\alpha)(H'\otimes H')\ket{\chi'_{\psi,\phi}}
\fine
where
\eq\label{chi'}
\ket{\chi'_{\psi,\phi}}=d_H\ket{H}_1\ket H_2+e^{\ii\Gamma}d_V\ket{V}_1\ket V_2
\fine
and the unitaries $P_\alpha$ and $H'$ are defined in \eqref{W}.
The gate $P_\alpha$ represents a phase shifter that adds a phase difference $\alpha=\frac{\psi}{2}+\pi$ between the states $\ket{V}$ 
and $\ket{H}$.
The gate $H'$ (similar to the Hadamard gate) performs the transformations $\ket{H}\rightarrow\frac{1}{\sqrt{2}}(\ket{H}-\ket{V})$ and 
$\ket{V}\rightarrow\frac{1}{\sqrt{2}}(\ket{H}+\ket{V})$
\footnote{Note that the transformation $H'$ is related to the usual Hadamard transformation $H$ by a unitary matrix, i.e. 
$H'=\sigma_zH$, where $\sigma_z$ is the usual Pauli matrix.}. 
These unitaries are attainable by simple linear optical elements as wave plates.

\begin{table} 
\begin{ruledtabular}
 \begin{tabular}{|c|cc|cc|cc|cc|cc|cc|}
             &  $\psi$ && $\phi$ && $\boldsymbol\alpha$ && 						$\boldsymbol d_H$      &&     $\boldsymbol d_V$ 		 && $\boldsymbol\Gamma$   &\\
 \hline
 $\ket{v_1}$ & $0$ && $0$ &&   $0$   &&&&&&&\\
 \cline{1-7}
 $\ket{v_2}$ & $\frac23\pi$ && $-\frac23\pi$ && $-\frac{2}{3}\pi$&& $\frac{\sqrt2-1}{\sqrt6}$ && $\frac{\sqrt2+1}{\sqrt6}$ &&     0   & \\
 \cline{1-7}
 $\ket{v_3}$ & $-\frac23\pi$ && $\frac23\pi$ && $\frac{2}{3}\pi$&&&&&&&\\
 \hline\hline
 $\ket{w_1}$ & $-\frac23\pi$ && $-\frac23\pi$ && $\frac{2}{3}\pi$&&&&&&&\\
 \cline{1-7}
 $\ket{w_2}$ & $\frac23\pi$ && $0$ &&  -$\frac{2}{3}\pi$&& $\sqrt{\frac{3+\sqrt2}{6}}$ && $\sqrt{\frac{3-\sqrt2}{6}}$ &&
  \text{arcsin}$\sqrt{\frac{6}{7}}$&\\
 \cline{1-7}
 $\ket{w_3}$ & $0$ && $\frac23\pi$ &&  $0$&&&&&&&\\
 \hline\hline
 $\ket{z_1}$ & $\frac23\pi$ && $\frac23\pi$ && $-\frac{2}{3}\pi$&&&&&&&\\
 \cline{1-7}
 $\ket{z_2}$ & $-\frac23\pi$ && $0$ && $\frac{2}{3}\pi$ && $\sqrt{\frac{3+\sqrt2}{6}}$ && $\sqrt{\frac{3-\sqrt2}{6}}$ &&    
  -\text{arcsin}$\sqrt{\frac{6}{7}}$&\\
 \cline{1-7}
 $\ket{z_3}$ & $0$ && $-\frac23\pi$ && $0$&&&&&&&\\
 \end{tabular}
  \end{ruledtabular}
   \caption{Theoretical values of $\alpha$, $d_H,d_V$ and $\Gamma$ for the states of the m.u. bases.}
   \label{table1}
 \end{table}

As said, we are interested in particular to generate three sets of m.u. bases. 
The (nine) vectors corresponding to the three bases sets, all expressed in the form of eq. \eqref{xi},
are explicitly given in the following:
\begin{align}
&\text{1)}\quad\bc
&\ket{v_1}=\frac1{\sqrt3}\left(\ket{HH}+\ket{VV}+\ket{\psi^+}\right)\\
&\ket{v_{2}}=\frac1{\sqrt3}\left(\ket{HH}+e^{\frac23\pi\ii}\ket{VV}+e^{-\frac23\pi\ii}\ket{\psi^+}\right)\\
&\ket{v_{3}}=\frac1{\sqrt3}\left(\ket{HH}+e^{-\frac23\pi\ii}\ket{VV}+e^{\frac23\pi\ii}\ket{\psi^+}\right)
\ec
\\
&\text{2)}\quad
\bc
&\ket{w_1}=\frac1{\sqrt3}\left(\ket{HH}+e^{-\frac23\pi\ii}\ket{VV}+e^{-\frac23\pi\ii}\ket{\psi^+}\right)\\
&\ket{w_{2}}=\frac1{\sqrt3}\left(\ket{HH}+e^{\frac23\pi\ii}\ket{VV}+\ket{\psi^+}\right)\\
&\ket{w_{3}}=\frac1{\sqrt3}\left(\ket{HH}+\ket{VV}+e^{\frac23\pi\ii}\ket{\psi^+}\right)
\ec
\\
&\text{3)}\quad
\bc
&\ket{z_1}=\frac1{\sqrt3}\left(\ket{HH}+e^{\frac23\pi\ii}\ket{VV}+e^{\frac23\pi\ii}\ket{\psi^+}\right)\\
&\ket{z_{2}}=\frac1{\sqrt3}\left(\ket{HH}+e^{-\frac23\pi\ii}\ket{VV}+\ket{\psi^+}\right)\\
&\ket{z_{3}}=\frac1{\sqrt3}\left(\ket{HH}+\ket{VV}+e^{-\frac23\pi\ii}\ket{\psi^+}\right)
\ec 
\end{align}
Note that in order to obtain a full set of m.u. bases, 
a fourth one, namely $\{\ket{HH},\ket{VV},\ket{\psi^+}\}$, must be considered \cite{86-woo-qua}. 

We give in Table \ref{table1} the explicit values of $\alpha$, $d_H,d_V$ and $\Gamma$ for all the states in the three m.u. bases. 
Detailed calculations are given in Appendix \ref{sec:calculus}.

\section{\label{sec:experiment}Experiment}
\begin{figure}[t]
	\centering
\includegraphics[scale=0.36]{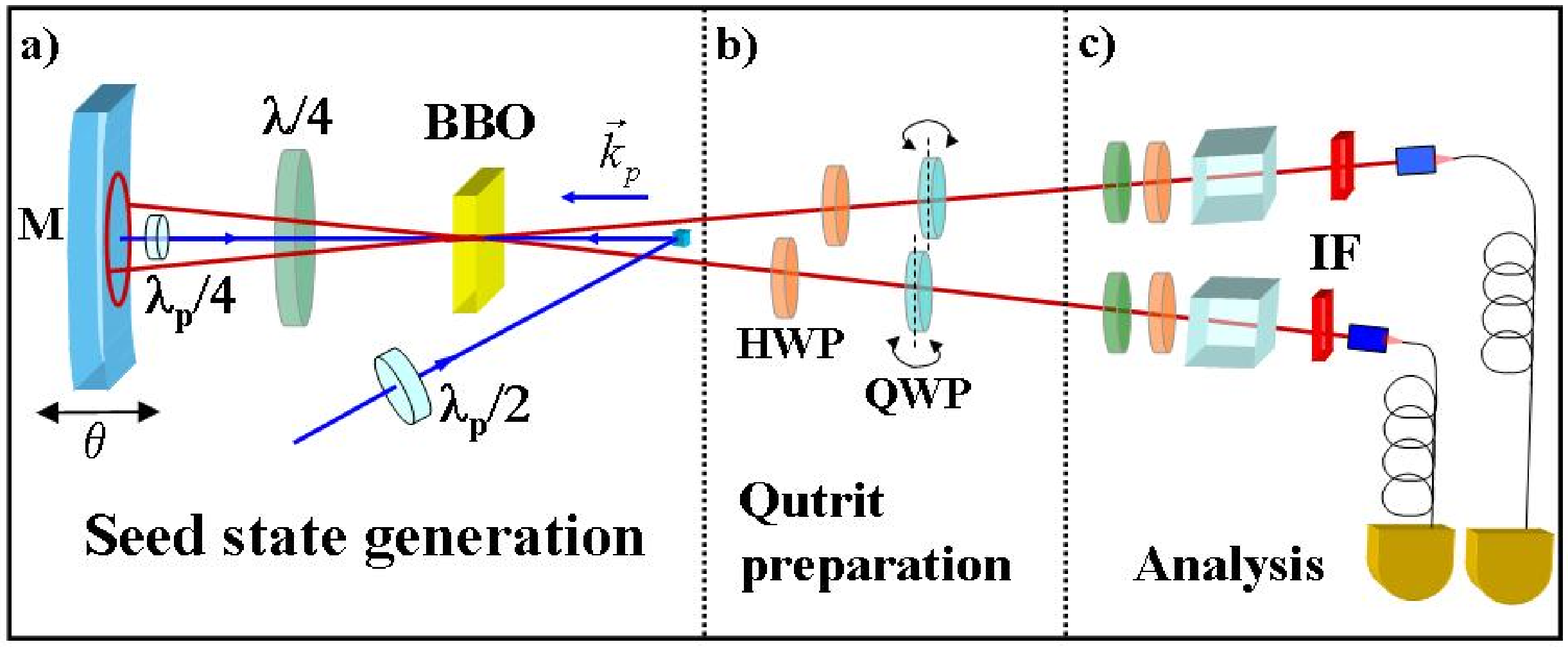}
	\caption{(Color online) Optical setup for generation and analysis of polarization qutrits. a) 
	The entanglement source is used to produce the seed state. 
The reciprocal weights of the $\ket{H}_1 \ket{H}_2$ and $\ket{V}_1 \ket{V}_2$ components are set by controlling the pump beam 
polarization in the first passage through BBO by the $\lambda_p/2$ half wave plate and in the second passage by the $\lambda_p/4$ quarter wave plate. 
b) The qutrit is encoded by applying the $H'\otimes H'$  transformation by two HWP plates and by 
proper phase shifts $P_\alpha\otimes P_\alpha$ performed by QWP plates.
c) Finally the state is characterized by polarization quantum state tomography.}
\label{fig:QT1}
\end{figure}
In this Section we explain how to implement the procedure described in Section \ref{sec:theory} and
show the obtained experimental results.
From eq. \eqref{xi'} and \eqref{chi'}, it follows that
all the states $\ket{\xi_{\psi,\phi}}$, expressed as \eqref{xi}, can be produced in four steps:
\begin{itemize}
\item[\bf I)] Choose $\phi$ and $\psi$ and generate the corresponding (non-maximally entangled) seed state $\ket{\chi_{\psi,\phi}}$. 
\item[{\bf II})] Change the relative phases between $\ket{H}_1\ket H_2$ and $\ket{V}_1\ket V_2$ in order to obtain $\ket{\chi'_{\psi,\phi}}$. 
\item[{\bf III})] Apply the gates $H'$ to each photon. This is performed by a half wave plate (HWP) whose axis is at -22.5$^\circ$ with respect 
to the horizontal direction.
\item[{\bf IV})] Apply the phase shifter $P_{\alpha}$ to each photon. This phase shift is realized by a birefringent medium, 
e.g. a quarter wave plate (QWP), with the optical axis oriented in the horizontal plane. 
The corresponding induced phase $\alpha$ is varied by rotating the plate along its vertical axix [cfr. Fig. \ref{fig:QT1}].
\end{itemize}
In the actual realization we performed step III) before step II). In this way 
the phase $\Gamma$ can be easily set by considering that the $H'\otimes H'$ gate transforms 
the seed $\ket{\chi'}$ in the following way:
\begin{multline}\label{Gamma}
 H'\otimes H'\ket{\chi'}=\\
 \frac{d_{H}+e^{\ii\Gamma}d_{V}}{2}(\ket{HH}+\ket{VV})-\frac{d_{H}-e^{\ii\Gamma}d_{V}}{\sqrt2}\ket{\psi^{+}}\,.
 \end{multline}
For fixed values of $d_H$ and $d_V$, the value of $\Gamma$ determines the relative weight of $\ket{HH}$ (or $\ket{VV}$) and $\ket{\psi^{+}}$. 
In this way the value of $\Gamma$ is chosen in order to make equal the two weights.

\subsection{\label{sec:source}Parametric Source}

Photon pairs are generated by a SPDC source whose detailed description is 
given in \cite{03-bar-det,04-bar-gen-PRL,04-cin-par}. 
It allows the efficient generation of the polarization entangled states $|\Phi _{\theta }\rangle=\frac{1}{\sqrt{2}}
(|H\rangle _{1}|H\rangle _{2}+e^{i\theta }|V\rangle _{1}|V\rangle _{2})$ by using a type I, $0.5mm$ thick, 
$\beta $-$BaB_{2}O_{4}$ (BBO) crystal. In the source, the entanglement arises from the superposition of the degenerate parametric emissions 
($\lambda=728nm$) of the crystal, excited in two opposite directions $\vec{k}_{p}$ and $-\vec{k}_{p}$ by a $V$-polarized 
Argon laser beam ($\lambda_p=364nm$). 
In the following we will refer to the emission excited in the direction $\vec{k}_p$ as the ``left'' emission 
(i.e. on the left of the BBO crystal in Fig. \ref{fig:QT1}), while the emission excited in the direction $-\vec{k}_p$ is the ``right'' one. 
The $H$-polarized photons belonging to the ``left'' emission are transformed $\ket{H}\rightarrow\ket{V}$ by a double passage
through a quarter wave plate ($\lambda/4$ in Fig. \ref{fig:QT1}).
Phase $\theta$ can be easily set by a micrometric translation of the spherical mirror M. 
Parametric radiation is coupled to two single mode fibers, achieving a coincidence level of $\sim1000$ sec$^{-1}$, 
over the 20nm bandwidth of two interference filters (IF, Fig. \ref{fig:QT1}).

By this source we can easily generate the states $\ket{HH}$, $\ket{VV}$ and $\ket{\psi^+}$.
The first two states are simply obtained by selecting only the right or left emission,
with fidelities $F_{\ket{HH}}=0.991\pm0.010$ and $F_{\ket{VV}}=0.960\pm0.008$. The state $\ket{\psi^+}$ can be generated
from the state $\ket{\Phi_0}$ by applying a HWP at $45^\circ$ on one photon, obtaining the fidelity $F_{\ket{\psi^+}}=0.966\pm0.008$.
{The fidelities of $\ket{HH}$ and $\ket{VV}$ are different mainly because of the non ideal behavior of 
the $\lambda/4$ waveplate. Indeed the operational wavelenght of all the waveplate adopted in our experiment is equal to 750nm.
As we shall see below, this feature partially affects the overall fidelities of the generated qutrits.

An other possible source of imperfection arises from the critical spatial matching between the right and left parametric emission.
This is overcome by the adoption of a thin crystal and single mode fibers.
Moreover by this scheme no temporal or spatial crystal walkoff is present with Type I phase matching.}

\subsection{\label{sec:seed}Seed state generation (Step I)}
The generation of non-maximally entangled states by the above described SPDC source was previously demonstrated in Ref. \cite{05-bar-tow}. 
The basic idea consists of tuning the polarization of the pump beam so that the nonlinear gain for the SPDC process can be varied. 
Indeed, if the pump beam is linearly polarized at an angle $\Theta_p$ with respect to the BBO optic axis, 
the SPDC probability is $p\propto\cos^2\Theta_p$. Therefore, by inserting 
a QWP intercepting only the pump beam
between the BBO and the mirror M ($\lambda_p/4$ in Fig. \ref{fig:QT1}), 
the right emission probability becomes lower and the seed state
\eq
\ket{\chi'}=d_H\ket{HH}+e^{\ii\Gamma}d_V\ket{VV},\quad d_H<d_V\,,
\fine
is generated. Phase $\Gamma$ is set by finely translating the spherical mirror, as said. 
On the other hand, seed states with higher $\ket{HH}$ component ($d_H>d_V$) can be generated by
inserting a further HWP before BBO ($\lambda_p/2$ in Fig. \ref{fig:QT1}). In this way, by changing the $\vec{k}_p$ pump polarization, we
lower the efficiency for the left emission. The $\lambda_p/4$ waveplate is used to rotate back the $-\vec{k}_p$ 
beam polarization to the vertical direction,
then raising the right emission. Then the states 
\eq
\ket{\chi'}=d_H\ket{HH}+e^{\ii\Gamma}d_V\ket{VV},\quad d_H>d_V\,,
\fine
are generated.
\begin{figure}[t]
	\includegraphics[scale=.37]{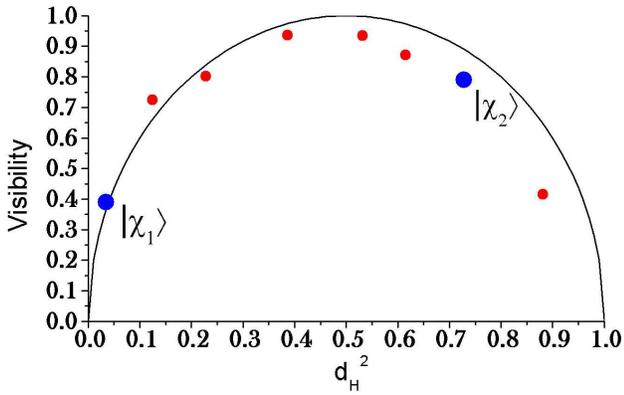}
	\caption{(Color online) Visibility (V) of non-maximally entangled state $\ket{\chi}$ vs. the $\ket{HH}$ weight $d_H^2$.
	The black line represents the theoretical curve, $V=2\sqrt{d^2_H(1-d^2_H)}$. Error bars are lower than the dimension of the point symbols.
}
	\label{fig:visibilita}
\end{figure}
 
For our experiment, two different seed states are needed (cfr. Table \ref{table1}), namely:
 \eq\label{2seeds}
 \begin{aligned}
 \ket{\chi_{1}}=&\frac{\sqrt{2}-1}{\sqrt6}\ket{HH}+\frac{\sqrt{2}+1}{\sqrt6}\ket{VV}\\
 {\simeq}&0.169\ket{HH}+0.986\ket{VV}\\
 \ket{\chi_{2}}=&\sqrt{\frac{3+\sqrt{2}}{6}}\ket{HH}+\sqrt{\frac{3-\sqrt{2}}{6}}\ket{VV}\\
  {\simeq}&0.858\ket{HH}+0.514\ket{VV}
 \end{aligned}
 \fine
The first seed state $\ket{\chi_{1}}$ is used for the first basis set $\{\ket{v_{a}}\}$, while the second seed state $\ket{\chi_{2}}$
 is used for the remaining two sets, namely $\{\ket{w_{a}}\}$ and $\{\ket{z_{a}}\}$. Note that the intrinsic difficulty to implement the 
 first state is due to the required unbalancement of the two contributions, $d_H^2/d_V^2\approx0.03$,
 almost comparable with the experimental uncertainties associated to each polarization contribution.

We show in Fig. \ref{fig:visibilita} the visibility $V=\frac{N_{max}-N_{min}}{N_{max}+N_{min}}$
of different non-maximally entangled state as a function of the  
probability $d_H^2$ of $\ket{HH}$. It is calculated by the coincidences of the two photons measured in the diagonal component 
$\frac{1}{\sqrt2}(\ket{H}+\ket{V})$ varying the phase $\Gamma$ from $0$ to $\pi$. 
$N_{max}$ ($N_{min}$) are the coincidence counts corresponding to $\Gamma=0$ ($\Gamma=\pi$).
The two blue points refer to the states $\ket{\chi_1}$ and $\ket{\chi_2}$. 
The points on the left ($d^2_H<0.5$) are closer to the theoretical curve probably because 
only the insertion of $\lambda_p/4$ is required for those states.
 \begin{figure}[t]
		\includegraphics[scale=.6]{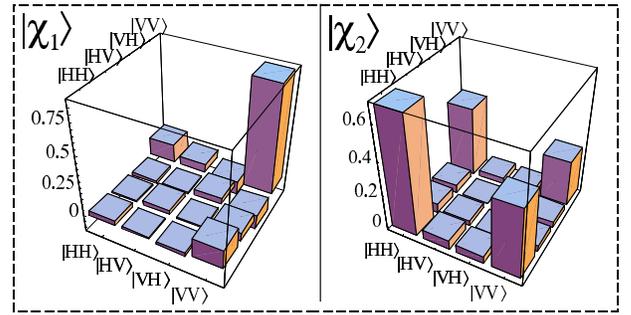}
		\caption{(Color online) Experimental quantum tomographies (real parts) of the 
		seed states $\ket{\chi_1}$ and $\ket{\chi_2}$ expressed in the $\{\ket{HH},\ket{HV},\ket{VH},\ket{VV}\}$ basis. 
		For the two states we obtain the purities $\mathcal P_{\ket{\chi_1}}=0.908\pm0.034$ and $\mathcal P_{\ket{\chi_2}}=0.930\pm0.036$.
		The imaginary components are negligible.
}
	\label{fig:tomo1}
\end{figure}

For a complete characterization of the two seed states \eqref{2seeds}, 
we performed a complete quantum tomography of the states, whose resulting diagrams are shown in Fig. \ref{fig:tomo1}. 
We used the ``Maximum Likelihood Estimation'' method described in \cite{01-jam-mea},
obtaining the fidelity  $F_1=0.912\pm0.010$ for $\ket{\chi_1}$ and $F_2=0.946\pm0.016$ for $\ket{\chi_2}$. 
We also measured the trace of the square of the experimental density matrix, i.e. the
purity of the generated states $\mathcal P_\rho=\text{Tr}[\rho^2]$. The results are given in the caption of Fig. \ref{fig:tomo1}.

\subsection{\label{sec:hada}$H'$ gate and $\Gamma$ phase setting (Steps II,III)}
\begin{figure}
	\centering
	\subfigure{\includegraphics[scale=.35]{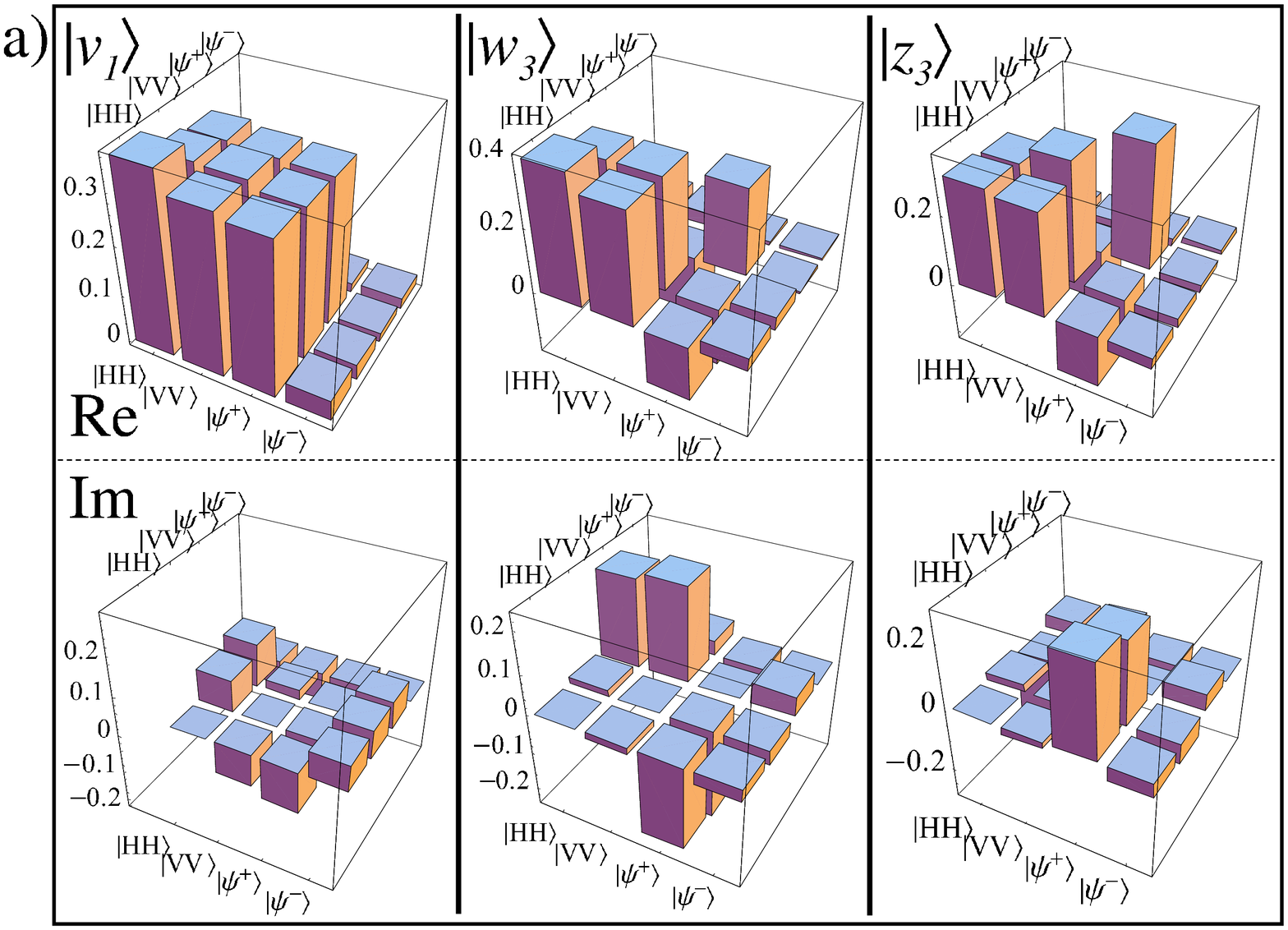}}
  \subfigure{\includegraphics[scale=.35]{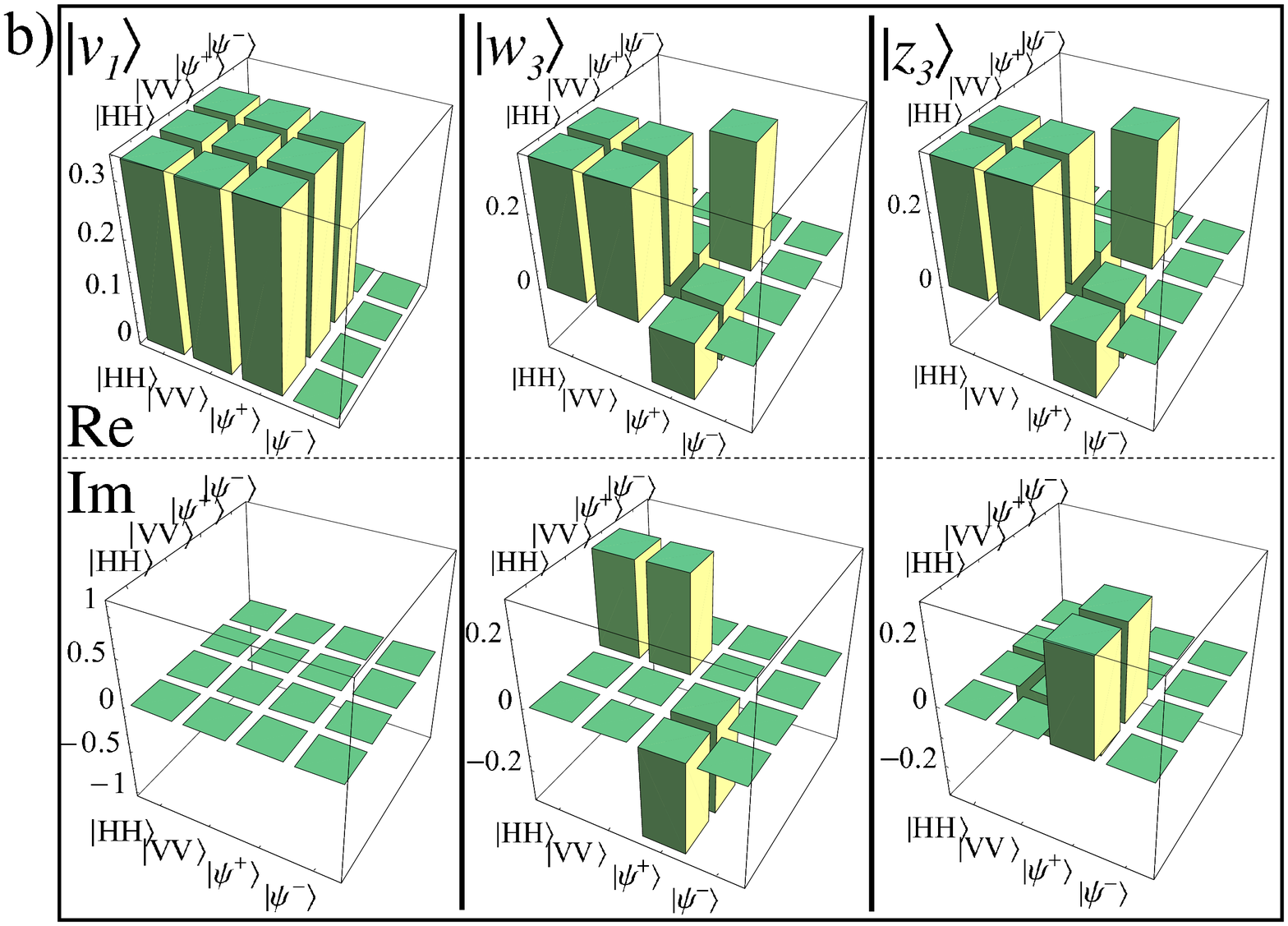}}
		\caption{(Color online) 
Experimental quantum tomography (a) and theoretical density matrices (b) of the 
states $\ket{v_1}$, $\ket{w_3}$ and $\ket{z_3}$. 
The upper pictures represent the real (Re) parts of the density matrices,
while the lower pictures represent the imaginary (Im) parts.
We measured the purities $\mathcal P_{\ket{v_1}}=0.974\pm0.030$, $\mathcal P_{\ket{w_3}}=0.904\pm0.033$, $\mathcal P_{\ket{z_3}}=0.895\pm0.028$.
\label{fig:qtrit4}
}
\end{figure}
The following steps for qutrit generation correspond to apply the $H'$
transformation [Fig. \ref{fig:QT1}] and the $\Gamma$ phase setting to each photon. As said, the $H'\otimes H'$ transformation 
is performed by the action of two HWP's oriented at $-22.5^\circ$ with respect to the vertical direction. 

Phase $\Gamma$ needed for the $\ket{\chi'}$ generation is set, as already said, after the insertion of 
the HWP wave plates that implement the unitary gate $H'\otimes H'$. The correct position is changed
by micrometric translation of the mirror M [see Fig. \ref{fig:QT1}] and fixed by observing 
that the count rate for $\ket{H}_1\ket{H}_2$ events doubles that of the $\ket{H}_1\ket{V}_2$ contribution.  

It is evident from Table \ref{table1} that the states $\ket{v_{1}},\ket{w_{3}}$ and $\ket{z_{3}}$ can be generated
by applying only the previous operations, i.e. without the need of inserting the phase gates $P_{\alpha}\otimes P_{\alpha}$. 
The corresponding experimental density matrices are shown in Fig. \ref{fig:qtrit4},
with fidelities $F_{\ket{v_1}}=0.949\pm0.010$,
$ F_{\ket{w_{3}}}=0.931\pm0.011$ and
$F_{\ket{z_{3}}}=0.932\pm0.010$.
Here and in the following we will use 
the basis $\{\ket{HH},\ket{VV},\ket{\psi^+},\ket{\psi^-}\}$ in order to have a better comparison with \eqref{xi}. 
These states are obtained by the insertion of two half wave-plate (HWP in Fig. \ref{fig:QT1}) and
correct phase $\Gamma$ setting (see Table \ref{table1}), as said.

\subsection{\label{sec:phase}Phase gate (Step IV)}
\begin{figure}
	\centering
		\subfigure{\includegraphics[scale=.45]{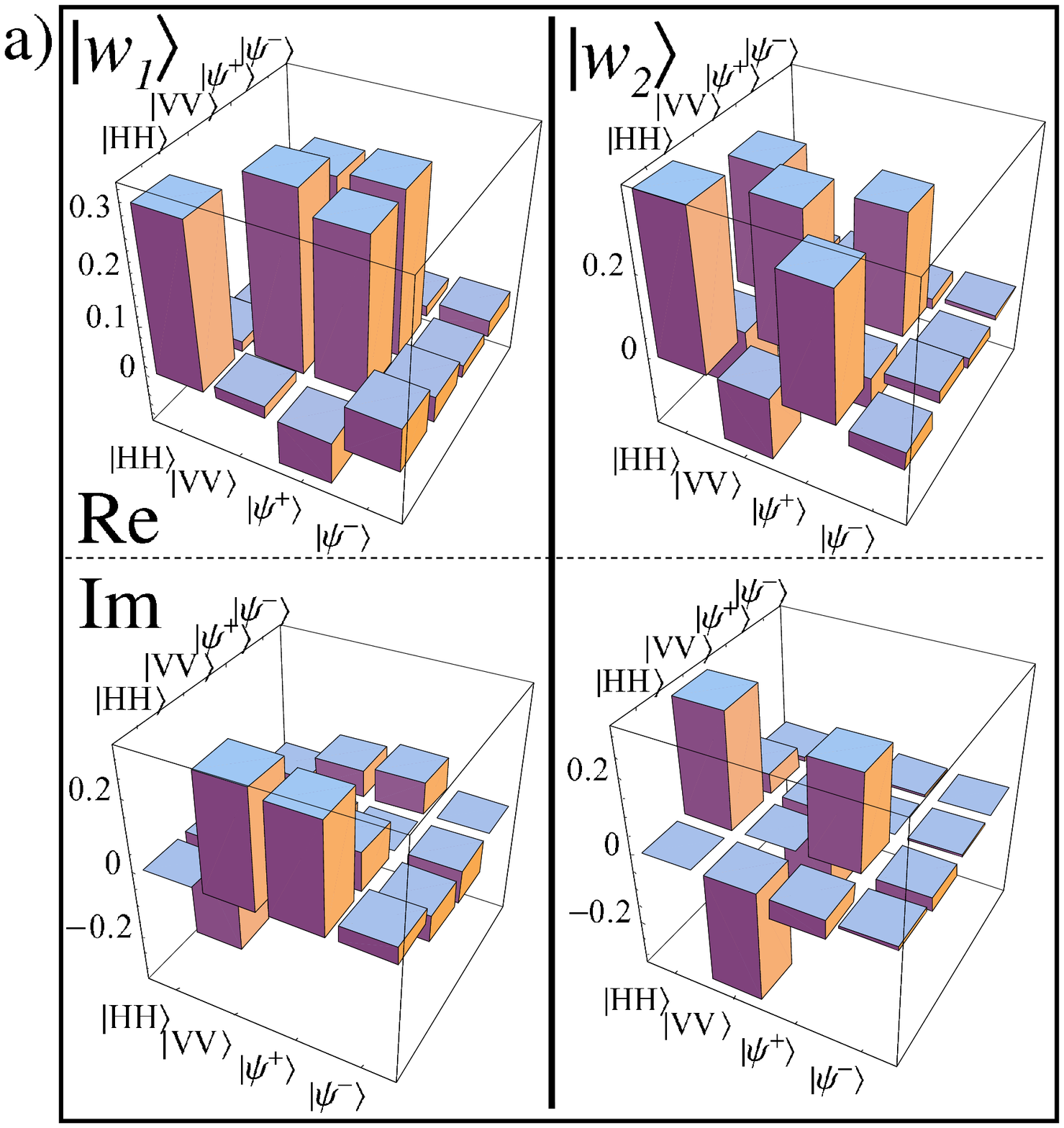}}
		\subfigure{\includegraphics[scale=.45]{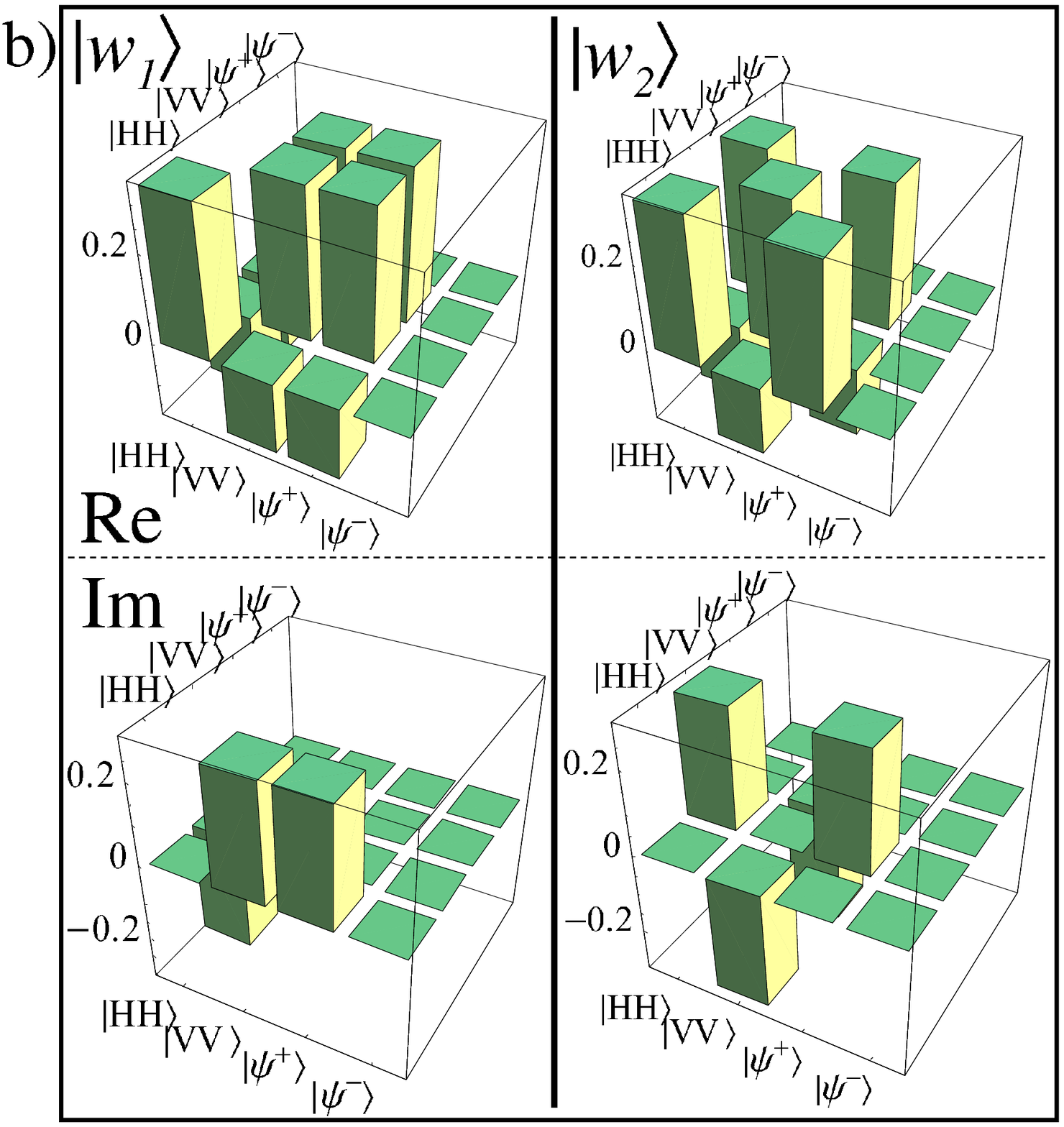}}
		\caption{(Color online) Experimental quantum tomography (a) and theoretical density matrices (b) of the 
states $\ket{w_1}$ and $\ket{w_2}$. 
We have the purities $\mathcal P_{\ket{w_1}}=0.969\pm0.030$, $\mathcal P_{\ket{w_2}}=0.918\pm0.024$.
}
	\label{fig:qtrit5}
\end{figure}
The implementation of the last gate of the protocol, namely the $P_\alpha\otimes P_\alpha$ operation,
is realized by inserting for each photon a QWP with vertical optical axis. It is mounted on a rotating stage which 
allows to tune the actual thickness.
In this way different phase shifts between the vertical and horizontal polarization components are achieved.
In Fig. \ref{fig:qtrit5} we show the two states $\ket{w_1}$ and $\ket{w_2}$ obtained by implementating the gate. 
The experimental fidelities are
$ F_{\ket{w_{1}}}=0.901\pm0.010$ and
$F_{\ket{w_{2}}}=0.939\pm0.009$.

\begin{figure}
	\centering
		\subfigure{\includegraphics[scale=.45]{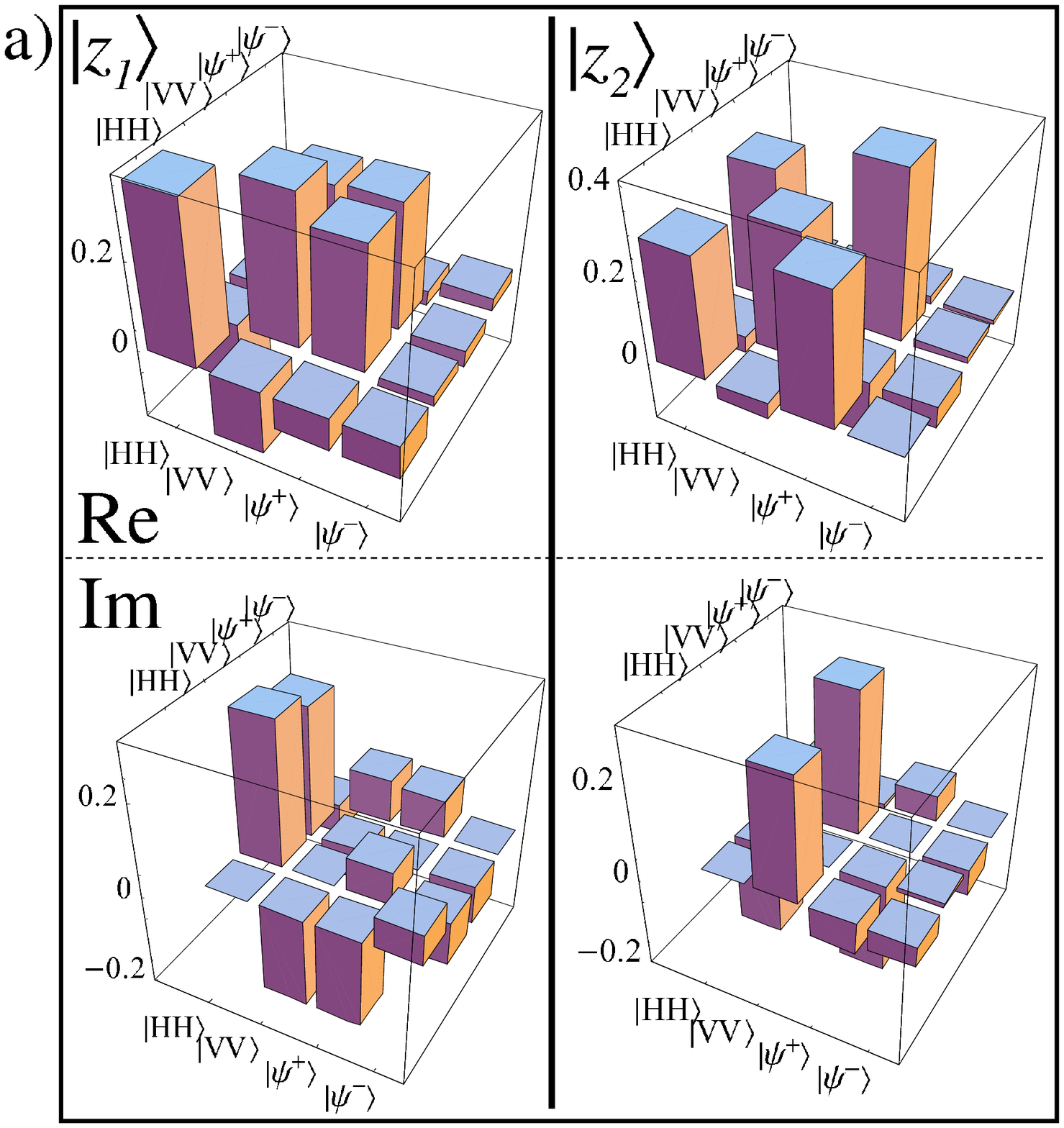}}
		\subfigure{\includegraphics[scale=.45]{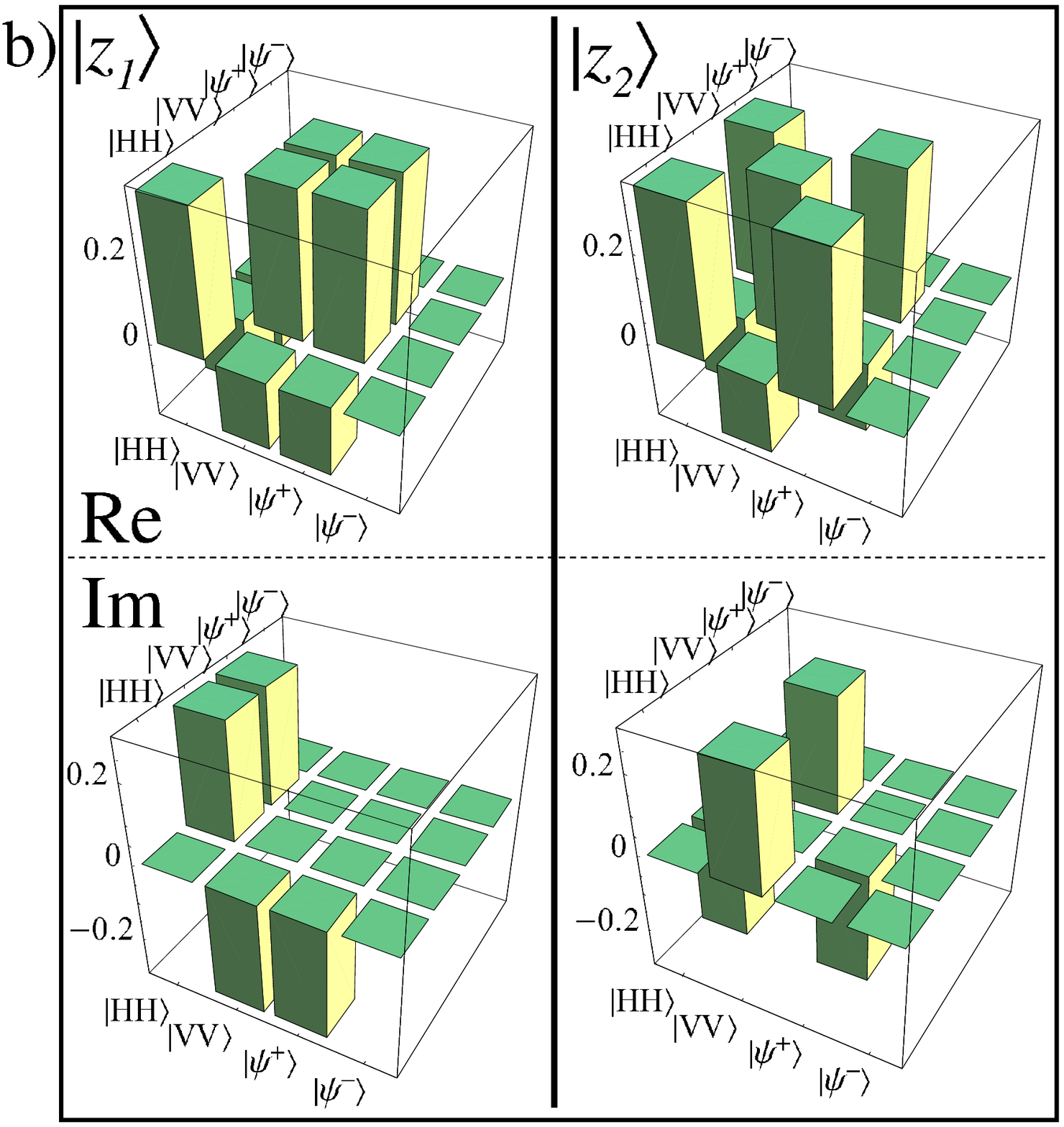}}
		\caption{(Color online) Experimental quantum tomography (a) and theoretical density matrices (b) of the 
states $\ket{z_1}$ and $\ket{z_2}$. 
We have the purities $\mathcal P_{\ket{z_1}}=0.931\pm0.028$, $\mathcal P_{\ket{z_2}}=0.937\pm0.032$.
}
	\label{fig:qtrit6}
\end{figure}

{We also generated the two remaining states of the $\ket{z_a}$ basis (see Fig. \ref{fig:qtrit6}).
The experimental fidelities are given by $F_{\ket{z_{1}}}=0.918\pm0.009$ and
$F_{\ket{z_{2}}}=0.933\pm0.009$. We did not actually generate the other two states 
$\ket{v_{2}}$ and $\ket{v_{3}}$ of the fourth basis, but we expect similar results for them. 
However it is well known that a qutrit-based quantum key distribution adopting only three mutually unbiased bases is more secure 
than qubit-based schemes \cite{00-bec-qua}.
Furthermore, it allows a higher transmission rate.
}

\section{Conclusions}
In this paper we have shown the experimental feasibility of the proposal given in \cite{05-dar-gen}
for the realization of polarization qutrit states.
The protocol starts from the generation of a two photon non-maximally entangled state and is based on
the application of two unitary transformations to each photon. Each relevant parameter of the qutrit states can be easily
tuned by this protocol. The experimental procedure can be described in four steps; 
we showed the experimental results corresponding to each step, demonstrating in this way the actual implementation of the procedure.
This methods is very powerful as demonstrated by the high coincidence rate and the high values of fidelities of the states.
Moreover the simplicity of this scheme could allow an easy experimental implementation of quantum security protocols.

\begin{acknowledgments}
We thank Massimiliano Sacchi and Mauro D'Ariano for useful discussions.
This work was supported by the PRIN 2005 ({\it New perspectives in entanglement and hyper-entanglement generation and manipulation})
of MIUR (Italy).

\end{acknowledgments}

\appendix	
\section{\label{sec:calculus}Calculation}
In this section we describe in detail how the transformations $U$ and $W$, which generate the state \eqref{xi}, are found.

Note that the state $\ket{\xi_{\phi,\psi}}$ can be also written as
\eq
\ket{\xi_{\phi,\psi}}=(\Lambda\otimes\openone)(\ket{H}_1\ket{H}_2+\ket{V}_1\ket{V}_2)\,,
\fine
where the matrix $\Lambda$ acting on photon 1 is written in the basis $\{\ket{H},\ket{V}\}$ as
\eq\label{Lambda}
\Lambda=\frac{1}{\sqrt3}
\begin{pmatrix}
1 & \frac1{\sqrt2}e^{\ii\phi}\\
\frac1{\sqrt2}e^{\ii\phi} & e^{\ii\psi}
\end{pmatrix}\,.
\fine
The unitaries $U$ and $W$ are then defined by the singular value decomposition of $\Lambda$:
\eq\label{UDW}
\Lambda=UDW^T\,,
\fine
where $D=\left(\begin{smallmatrix}d_H&0\\0&d_V\end{smallmatrix}\right)$
is the diagonal matrix with eigenvalues equal to the positive square roots of the eigenvalues of $\Lambda^\dag\Lambda$.
In the previous equation $W^T$ means the transpose in the basis $\{\ket{H},\ket{V}\}$.

From \eqref{Lambda} it follows that
\eq
\begin{aligned}
\ket{\xi_{\phi,\psi}}=&(UDW^T\otimes\openone)(\ket{HH}+\ket{VV})=\\
=&(UD\otimes W)(\ket{HH}+\ket{VV})=\\
=&(U\otimes W)(d_H\ket{HH}+d_V\ket{VV})
\end{aligned}
\fine

Let's now find the matrices $U$ and $W$ in a explicit way.
By virtue of decomposition \eqref{UDW},
the unitary transformation $W^T$ is the matrix that diagonalizes $\Lambda^\dag\Lambda$:
\eq\label{diagonalize}
\begin{aligned}
\Lambda^\dag\Lambda&=
(W^T)^\dag XW^T
\end{aligned}
\fine
where 
\eq
X=\begin{pmatrix}
|\z_+|^2 & 0\\
0 & |\z_-|^2
\end{pmatrix}
\,\Rightarrow\,
D=\begin{pmatrix}
|\z_+| & 0\\
0 & |\z_-|
\end{pmatrix}
\fine
and
$\z_\pm$ are defined in \eqref{x+-}. The explicit values of the elements of $D$ are
\eq
\begin{aligned}
d_H=|\z_+|=\sqrt{\frac12+\frac{\sqrt2}3\cos(\phi-\frac\psi2)}\\
d_V=|\z_-|=\sqrt{\frac12-\frac{\sqrt2}3\cos(\phi-\frac\psi2)}
\end{aligned}
\fine
From \eqref{diagonalize} we find the unitary $W$ as
\eq\label{W app}
W=\frac1{\sqrt2}
\begin{pmatrix}
1 & 1\\
e^{\ii\frac\psi2} & -e^{\ii\frac\psi2}
\end{pmatrix}
\fine
Note that the matrices $U$ and $W$ are defined up to the following transformation
\eq
\begin{cases}
U\rightarrow UZ\\
W\rightarrow WZ^\dag
\end{cases},
\quad\text{where }
Z=\begin{pmatrix}
e^{\ii z_1} & 0\\
0 & e^{\ii z_2}
\end{pmatrix}
\fine
and $e^{\ii z_{1,2}}$ correspond to the global phases chosen for the eigenvectors of $\Lambda^\dag\Lambda$.
Equation \eqref{W app} is then only one of the infinite solutions for $W$.

The matrix $U$ is easily found from \eqref{UDW}
\eq
\begin{aligned}
U=\Lambda (W^T)^\dag D^{-1}=\frac{\z_+}{|\z_+|}
W
\begin{pmatrix}
1& 0\\
0 & e^{\ii\Gamma}
\end{pmatrix}
\end{aligned}
\fine
and $\Gamma$ is defined in \eqref{U}:
\eq
\begin{aligned}
&\Gamma=\text{arg}\left(\frac{\z_-}{\z_+}\right)=\beta-\gamma\\
&\beta=\text{arg}[\sqrt2-e^{\ii(\phi-\frac\psi2)}]\\
&\gamma=\text{arg}[\sqrt2+e^{\ii(\phi-\frac\psi2)}]
\end{aligned}
\fine
We note that the previous expression of $U$ differs from equation \eqref{U} for the phase $\frac{\z_+}{|\z_+|}$.
However this is only a global phase and can be discarded.

Let's now find a more explicit expression of $\Gamma$. From the 
previous equation we have
\eq
\bc
&\sin\gamma=\frac{\sin(\phi-\frac{\psi}{2})}{\sqrt{3+\sqrt8\cos(\phi-\frac{\psi}{2})}}\\
&\cos\gamma=\frac{\sqrt2+\cos(\phi-\frac{\psi}{2})}{\sqrt{3+\sqrt8\cos(\phi-\frac{\psi}{2})}}
\ec
\fine 
and 
\eq
\bc
&\sin\beta=-\frac{\sin(\phi-\frac{\psi}{2})}{\sqrt{3-\sqrt8\cos(\phi-\frac{\psi}{2})}}\\
&\cos\beta=\frac{\sqrt2-\cos(\phi-\frac{\psi}{2})}{\sqrt{3-\sqrt8\cos(\phi-\frac{\psi}{2})}}
\ec
\fine
The required expression for $\Gamma$ is then
\eq
\bc
&\sin\Gamma=-\frac{2\sqrt2\sin(\phi-\frac{\psi}{2})}{\sqrt{9-8\cos^2(\phi-\frac{\psi}{2})}}\\
&\cos\Gamma=\frac{1}{\sqrt{9-8\cos^2(\phi-\frac{\psi}{2})}}\\
\ec
\fine

\end{document}